\documentclass[12pt]{article}

\usepackage{epsfig}

\begin{document}

\title{
Doorway states for one-nucleon transfer reactions
as a test for current approaches to nuclei\thanks{Devoted to the memory
of \bf Sergey Fayans.}}
\author{B.L. Birbrair \& V.I.  Ryazanov\\ PNPI, Gatchina 188300, Russia
\thanks{birbrair@thd.pnpi.spb.ru}\\ } \date{} \maketitle

\begin{abstract}
The naturalness concept of the effective field theory is not confirmed.
\end{abstract}

\section{Introduction}

The doorway states for one-nucleon transfer reactions are eigenstates
of nucleon in the static field of nucleus which is the convolution of
the free-space nucleon-nucleon forces with the nucleon density
distributions in nucleus \cite{1}. They are model-independent
quantities because (a)~the free-space $NN$ forces are independent of
nuclear medium effects and (b)~the nucleon density distributions are
deduced from the electron--nucleus \cite{2} and proton--nucleus
\cite{3} elastic scattering data. The corresponding eigenvalue problem
is that of nucleon in central field which can be solved with any
desired accuracy. For these reasons the doorway states can be used as a
very trustworthy test for current nuclear models.

Let us discuss some results of Ref.\cite{1} from this point of view.

1. The nuclear relativity within the Walecka \cite{4} model is
confirmed to be actually existing phenomenon.

2. The dominant contribution to the isovector part of the static field
arises from the many-particle $NN$ forces because the $\rho$ meson
(vector--isovector) and $\delta$ meson (scalar--isovector) fields
arising from the two-particle forces nearly cancel each other. At the
same time the isovector nuclear potential is exclusively of the $\rho$
meson origin within the quantum hadrodynamics \cite{5,6}. The reason
for this wrong QHD result is the neglect of the $\delta$ meson field in
spite of the fact that both the $\rho$ and $\delta$ meson exchanges are
taken into account in the two-particle $NN$ forces \cite{7,8,9}.

3. The contributions from the two-particle, three-particle and
four-particle forces to the isoscalar part of the static field are
found to be $U_2\approx-80\,$MeV, $U_3\approx+96\,$MeV,
$U_4\approx-104\,$MeV. This is in conflict with such leading principles
of the effective field theory as the naive dimensional analysis and the
naturalness \cite{10,11}. Indeed, the values of the forces which are
estimated according to the above principles are \cite{12}
\begin{equation}
V_2\approx30\mbox{ MeV }, \quad V_3\approx\frac{V^2_2}m\approx1\mbox{
MeV }, \quad V_4\approx\frac{V^3_2}{m^2}\approx0.03\mbox{ MeV },
\end{equation}
($m$ is the mass of nucleon) and therefore the expected relation
between $U_2$, $U_3$ and $U_4$ is
$|U_2|:|U_3|:|U_4|\approx1:10^{-\frac32}:10^{-3}$. In fact it is
$1:1.2:1.3$ thus suggesting that there is something wrong
with the naturalness. Discussion of this point is continued in the next
Section.

\section{Nonlinearity as a source of many-particle forces}

As discussed in Ref.\cite{13} the isoscalar part of the static field may
contain contributions from higher (five-particle, six-particle, etc.)
many-particle forces. They could be taken into account by increasing
the number of terms in the power series expansion
\begin{equation}
U_{st}(r)\ =\ \sum^\infty_{n=2}\ a_n\rho^n(r)
\end{equation}
for the static field
\cite{1} ($\rho(r)$ is the nucleon density distribution) thus
introducing an indeterminate number of additional adjustable
parameters. Instead we use the fact that ultimately the underlying
reason for many-particle forces is the nonlinearity of strong
interaction. We introduce an auxiliary scalar-isoscalar field $\phi$
with the Lagrangian density
\begin{eqnarray} && L\ =\ \frac12\
\partial_\mu\phi\partial^\mu\phi-U(\phi)-g\bar\psi\psi\phi\\
&& U(\phi)\ =\ \frac12\ \Lambda^2\phi^2+\frac13\ \lambda_3\phi^3
+\frac14\ \lambda_4\phi^4\ ,
\end{eqnarray}
thus obeying the following
equation:
\begin{equation}
\Lambda^2\phi+\lambda_3\phi^2+\lambda_4\phi^3\ =\
-g\rho_s+ \Delta\phi\ ,
\end{equation}
$\rho_s(r)=\langle
A_0|\bar\psi(r)\psi(r)|A_0\rangle$ is the nuclear scalar density. As
discussed in Ref.\cite{1} the many-particle forces of all ranks are
taken into account in this way.

The field $\phi$ contains the "two-particle" component $\phi_2$ obeying
the equation
\begin{equation}
\Lambda^2\phi_2\ =\ -g\rho_s+\Delta\phi_2\ .
\end{equation}
This component must be eliminated because the "two-particle"
contribution to the static field is determined by the free-space
two-particle forces \cite{8,9}. So the many-particle contribution to
the scalar--isoscalar field is
\begin{equation}
W(r)\ =\ g(\phi(r)-\phi_2(r))\ .
\end{equation}

Let us analyse this expression disregarding for a moment the Laplace
terms of Eqs. (5) and (6) which are responsible for the finite range of
the forces, although these terms are included in the actual
calculations. But as demonstrated in Ref.\cite{13} they are of little
importance thus not affecting the results of the below analysis. As
demonstrated in Ref.\cite{1} the radial dependence of $W(r)$ has the
form which is schematically shown in Fig.1. As seen from the figure it
is negative at $r<r_1$, positive at $r>r_1$ with a maximum $W_m$ in
this region, and vanishing at $r=r_1$. Without the Laplace terms
\begin{equation}
W(r)\ =\ -\ \frac{g\lambda_4}{\Lambda^2}\ \phi^2(r)\left(
\frac{\lambda_3}{\lambda_4}+\phi(r)\right).
\end{equation}
So
\begin{equation}
\phi(r_1)\ =\ -\ \frac{\lambda_3}{\lambda_4}\ .
\end{equation}
But as follows from (6) and (7)
\begin{equation}
\phi(r_1)\ =\ -\ \frac g{\Lambda^2}\ \rho_s(r_1)
\end{equation}
and therefore
\begin{equation}
\frac{\lambda_3}{\lambda_4}\ =\ \frac{g\rho_1}{\Lambda^2}\ ,
\end{equation}
where $\rho_1=\rho_s(r_1)$. Let us introduce the dimensionless
quantities $y(r)$ and $y_2(r)$,
\begin{equation}
\phi(r)\ =\ -\ \frac g{\Lambda^2}\ \rho_1y(r)\ , \quad \phi_2(r)\ =\ -\
\frac g{\Lambda^2}\ \rho_1y_2(r)\ .
\end{equation}
In these units
\begin{equation}
W(r)\ =\ -\ \frac{g^4\rho^3_1}{\Lambda^8}\ y^2(r)(1-y(r))\lambda_4\ .
\end{equation}
The maximum $W_m$ occurs at $y=2/3$, so
\begin{equation}
\lambda_4=-\frac{27\Lambda^8W_m}{4g^4\rho^3_1}\ , \qquad
\lambda_3=-\frac{27\Lambda^6W_m}{4g^3\rho^2_1}\ .
\end{equation}
The parameter $\lambda_4$ is negative since $W_m>0$, see the figure.
The parameter $\lambda_3$ is negative  too provided the coupling
constant $g$ is positive. Actually the sign of $g$ is insignificant
since the physical field is $g\phi$ thus being expressed through $g^2$.

The potential energy of the scalar field is also of the form (4) within
the relativistic mean-field approach \cite{14,15}, the parameters
$\lambda_3$ and $\lambda_4$ being negative too. In this way the sign of
the RMF parameters is confirmed. It should be mentioned that the values
of the RMF parameters are determined from the experimental data which
include the important correlation effects (binding energies, density
distributions, low-energy spectra etc.), and therefore they are
model-dependent (the model-independent treatment of the
correlations does not exist). In contrast, our parameters are
determined from the doorway state energies thus being
model-independent.

In terms of the $y$ and $y_2$ quantities the contribution to the
scalar--isoscalar field from the many-particle forces is
\begin{eqnarray}
&& W(r)\ =\ -\frac{9xW_m}4(y(r)-y_2(r))+\frac12\beta\left(\rho_s^-
(r)\right)^2\\
&& x\ =\ \frac{4g^2\rho_1}{9\Lambda^2W_m}\ ,\ \quad \rho^-_s(r)\ =\
\rho_{sn}(r)-\rho_{sp}(r)\ .
\end{eqnarray}
The second term in the rhs of (15) arises from the symmetry energy. The
quantities $y(r)$ and $y_2(r)$ obey the equations
\begin{eqnarray}
&& y(r)+\frac3xy^2(r)(1-y(r))\ =\ \frac{\rho_s(r)}{\rho_1}
+\frac1{\Lambda^2}\ \Delta y(r)\\
&& y_2(r)\ =\ \frac{\rho_s(r)}{\rho_1}+\frac1{\Lambda^2}\ \Delta
y_2(r)\ .
\end{eqnarray}

The details of calculations will be described in the forthcoming
publication. The resulting values of the parameters are found to be
\begin{equation}
\rho_1=0.146\mbox{ fm}^{-3}, \quad W_m=11.393\mbox{ MeV }, \quad
x=16.004\ , \quad \Lambda=986.64\mbox{ MeV }, \quad \beta=5.583\mbox{
fm}^5.
\end{equation}

The NDA prescription \cite{10,11} for the scalar field potential energy
is \cite{16}
\begin{equation}
U(\phi)\ =\ f^2_\pi\Lambda^2\sum^\infty_{n=2}\ \frac{\kappa_n}{n~!}
\left(\frac\phi{f_\pi}\right)^n,
\end{equation}
where $f_\pi=93$ MeV. According to the concept of naturalness all the
coefficients $\kappa_n$ must be of order of unity. Comparison between
(20) and (4) together with (14) and (16) gives
\begin{equation}
\kappa_2=1, \quad \kappa_3=\frac{2f_\pi}{\Lambda^2}\lambda_3=
-4\Lambda\frac{f_\pi}{x\rho_1}\left(\frac{\rho_1}{xW_m}\right)^{1/2},
\quad \kappa_4=\frac{6f^2_\pi}{\Lambda^2}\lambda_4=-8\Lambda^2
\frac{f^2_\pi}{x^2\rho_1W_m}\ .
\end{equation}
As follows from the values (19) of the parameters $\kappa_3=-1.6$,
$\kappa_4=-20.5$, the concept of naturalness thus being not confirmed.

As demonstrated by the  calculations for the few-nucleon systems the
effect of many-particle forces is relatively small \cite{12}. This
result is confirmed but the underlying physical reason is different
from that provided by the effective field theory. According to the
latter it is decrease of the strength with increasing rank of the
force, see Eq.(1). As follows from above this scenario does not hold:
the actual reason is the cancellation of the contributions from
many-particle forces of different ranks (the physics is believed to be
the same for complex nuclei and few-nucleon systems).

\newpage
\begin{figure}
\centerline{
\epsfxsize=9.5cm\epsfbox{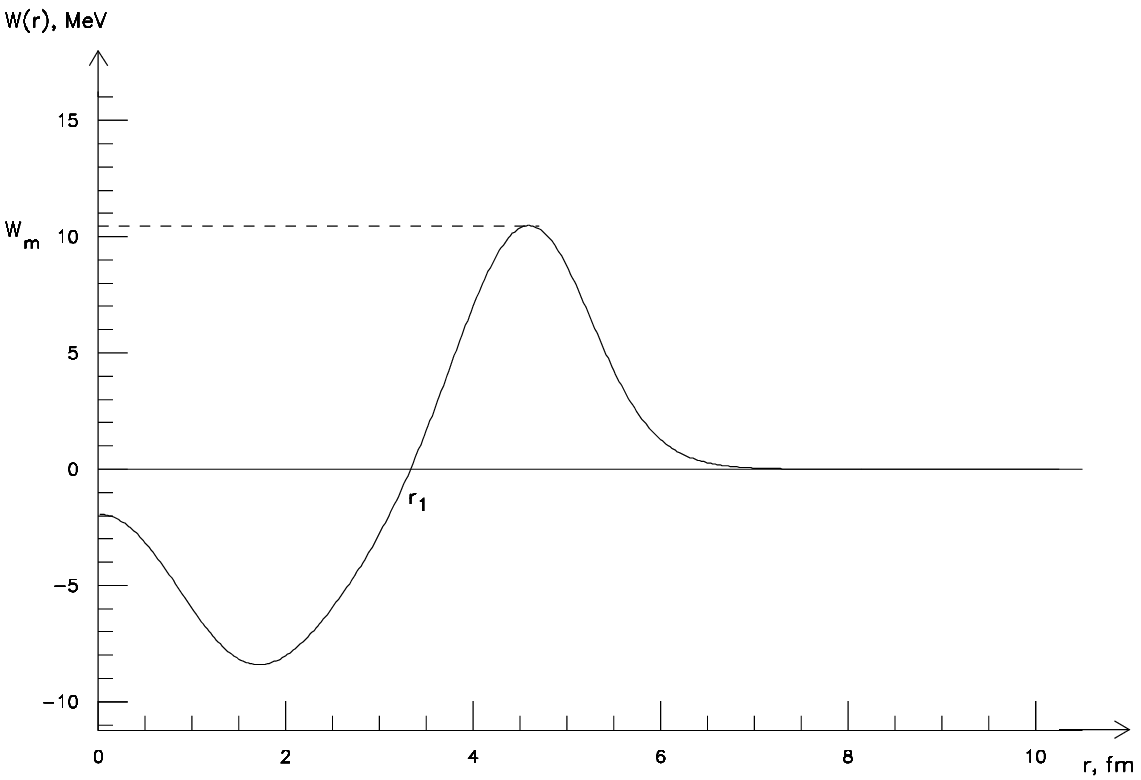}}

\vspace*{3cm}
\caption{Many-particle contribution to the static field.
}
\end{figure}

\end{document}